\documentclass[10pt,twocolumns]{article}

\usepackage{amssymb,amsmath,latexsym,amsthm,amsfonts}
\usepackage{graphicx}
\usepackage{fancyhdr,enumerate,scrtime}
\usepackage{fullpage}
\usepackage{color}
\usepackage{bbm}
\usepackage{mathrsfs}
\usepackage{ifthen}
\usepackage{natbib}

\bibpunct{(}{)}{,}{a}{}{,}

\newtheorem{example*}{Example}

\title{Monte Carlo Methods in Statistics}

\author{
{\sc Christian Robert}\thanks{Professor of Statistics, CEREMADE, Universit\'e Paris
Dauphine, 75785 Paris cedex 16, France. Supported by the Agence Nationale de la Recherche 
(ANR, 212, rue de Bercy 75012 Paris) through the 2009-2012 project ANR-08-BLAN-0218 {\sf Big'MC}. 
Email: \texttt{xian@ceremade.dauphine.fr}. The author is grateful to Jean-Michel Marin for helpful
comments.}\\ 
Universit\'e Paris Dauphine and CREST, INSEE
}

\begin{document}

\twocolumn
\maketitle

Monte Carlo methods are now an essential part of the statistician's toolbox, 
to the point of being more familiar to graduate students than the measure
theoretic notions upon which they are based! We recall in this note some of
the advances made in the design of Monte Carlo techniques towards their use
in Statistics, referring to \cite{robert:casella:2004,robert:casella:2010} for an
in-depth coverage. 

\section*{The basic Monte Carlo principle and its extensions}\label{sec:MCdebase}

\begin{figure}
\centerline{\includegraphics[width=.4\textwidth]{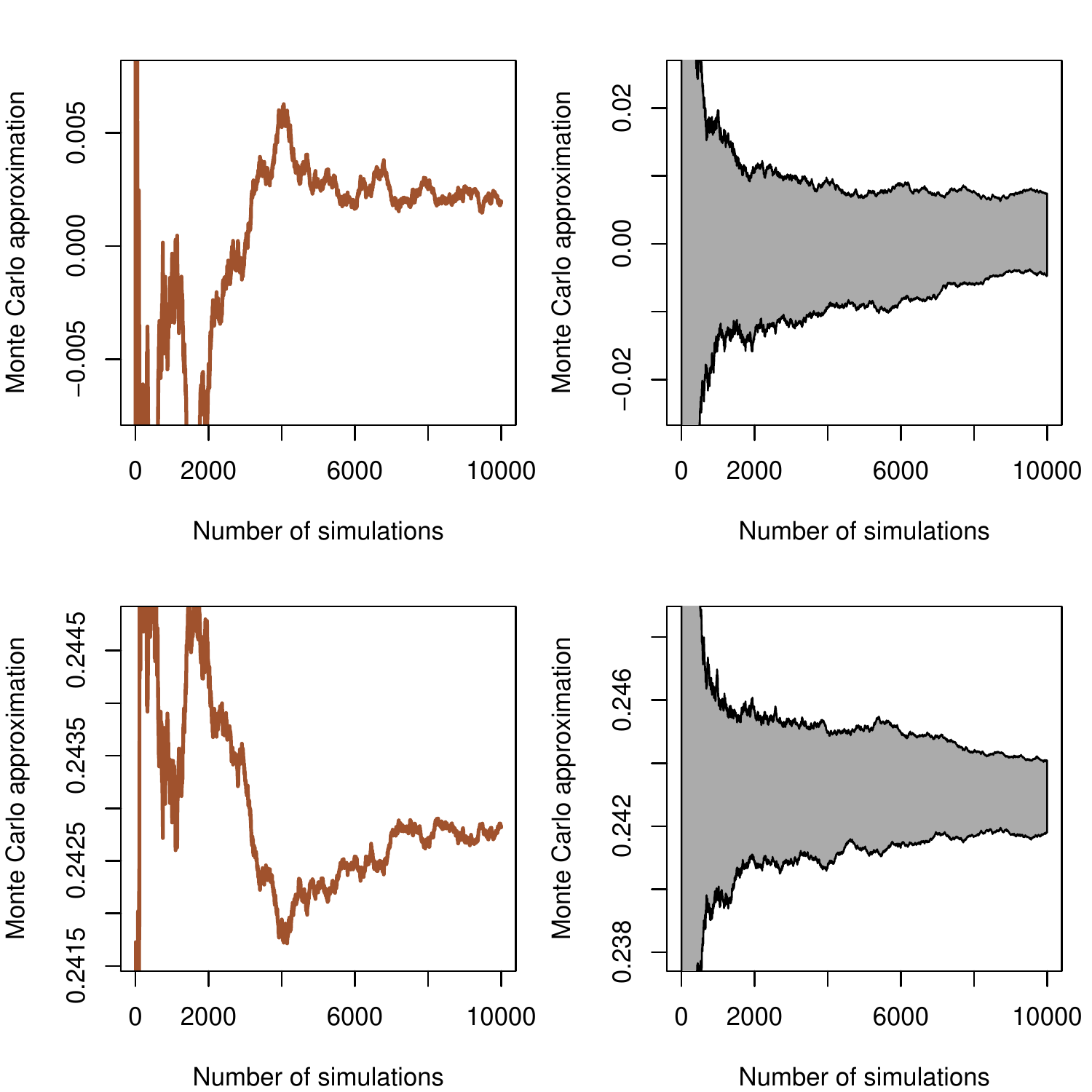}}
\caption{\label{fig:baseMC}
Monte Carlo evaluation \eqref{eq:MCdebase} of the expectation $\mathbb{E}[X^3/(1+X^2+X^4)]$ as a function of
the number of simulation when $X\sim\mathcal{N}(\mu,1)$ using {\em (left)} one simulation run 
and {\em (right)} 100 independent runs for {\em (top)} $\mu=0$ and {\em (bottom)} $\mu=2.5$.
}
\end{figure}

The most appealing feature of Monte Carlo methods [for a statistician] is that they rely on sampling
and on probability notions, which are the bread and butter of our profession. Indeed, the foundation of Monte
Carlo approximations is identical to the validation of empirical moment estimators in that the average
\begin{equation}\label{eq:MCdebase}
\frac{1}{T}\,\sum_{t=1}^T h(x_t)\,,\qquad x_t \sim f(x)\,,
\end{equation}
is converging to the expectation $\mathbb{E}_f[h(X)]$ when $T$ goes to infinity. Furthermore, the precision of this
approximation is exactly of the same kind as the precision of a statistical estimate, in that it usually evolves
as $\text{O}(\sqrt{T})$. Therefore, once a sample $x_1,\ldots,x_T$ is produced according to a distribution density
$f$, all standard statistical tools, including bootstrap, apply to this sample (with the further appeal that more
data points can be produced if deemed necessary). As illustrated by Figure \ref{fig:baseMC}, the variability due to
a single Monte Carlo experiment must be accounted for, when drawing conclusions about its output and evaluations of
the overall variability of the sequence of approximations are provided in \cite{kendall:marin:robert:2007}.
But the ease with which such methods are analysed and the systematic resort to statistical intuition explain in part
why Monte Carlo methods are privileged over numerical methods.


The representation of integrals as expectations $\mathbb{E}_f[h(X)]$ is far from unique and there
exist therefore many possible approaches to the above approximation. This range of choices corresponds
to the importance sampling strategies \citep{rubinstein:1981} in Monte Carlo, based on the obvious identity
$$
\mathbb{E}_f[h(X)] = \mathbb{E}_g[h(X) f(X)/ g(X)]
$$
provided the support of the density $g$ includes the support of $f$. Some choices of $g$ may however lead to appallingly
poor performances of the resulting Monte Carlo estimates, in that the variance of the resulting empirical
average may be infinite, a danger worth highlighting since often neglected while having a major impact
on the quality of the approximations. From a statistical perspective, there exist some natural choices for the importance
function $g$, based on Fisher information and analytical approximations to the likelihood function like the Laplace approximation
\citep{rue:martino:chopin:2008}, even though it is more robust to replace the normal distribution in the
Laplace approximation with a $t$ distribution. The special case of Bayes factors \citep{robert:casella:2004}
$$
B_{01}(x) =   { \displaystyle{\int_{\Theta} f(x|\theta) \pi_0(\theta) \text{d}\theta} }\bigg/ {
       \displaystyle{\int_{\Theta} f(x|\theta) \pi_1(\theta) \text{d}\theta} }\,,
$$
which drive Bayesian testing and model choice,
and of their approximation has led to a specific class of importance sampling techniques known as {\em bridge 
sampling} \citep{chen:shao:ibrahim:2000} where the optimal importance function is made of a mixture of the
posterior distributions corresponding to both models (assuming both parameter spaces can be mapped into the same
$\Theta$). We want to stress here that an alternative approximation
of marginal likelihoods relying on the use of {\em harmonic means} \citep{gelfand:dey:1994,newton:raftery:1994} and 
of direct simulations from a posterior density has repeatedly been used in the literature, despite often suffering 
from infinite variance (and thus numerical instability). Another potentially very efficient approximation of Bayes
factors is provided by Chib's (\citeyear{chib:1995}) representation, based on parametric estimates to the posterior
distribution.

\section*{MCMC methods}
Markov chain Monte Carlo (MCMC) methods have been proposed many years 
\citep{metropolis:rosenbluth:rosenbluth:teller:teller:1953} before their impact in Statistics was truly felt.
However, once \cite{Gelfand:Smith:1990} stressed the ultimate feasibility of producing a Markov chain with a given
stationary distribution $f$, either via a Gibbs sampler that simulates each conditional distribution of $f$ in its turn,
or via a Metropolis--Hastings algorithm based on a proposal $q(y|x)$ with acceptance probability [for a move from $x$ to $y$]
$$
\min\left\{1,{f(y)q(x|y)}\big/{f(x)q(y|x)}\right\}\,,
$$
then the spectrum of manageable models grew immensely and almost instantaneously. 

Due to parallel developments at the time on
graphical and hierarchical Bayesian models, like generalised linear mixed models
\citep{zeger:karim:1991}, the wealth of multivariate models with available conditional
distributions (and hence the potential of implementing the Gibbs sampler) was far from negligible, especially when 
the availability of latent variables became quasi universal
due to the slice sampling representations \citep{damien:wakefield:walker:1999,neal:2003}. (Although the adoption of Gibbs
samplers has primarily taken place within Bayesian statistics, there is nothing that prevents an artificial augmentation of the data
through such techniques.) 

\begin{figure}
\centerline{\includegraphics[width=.45\textwidth]{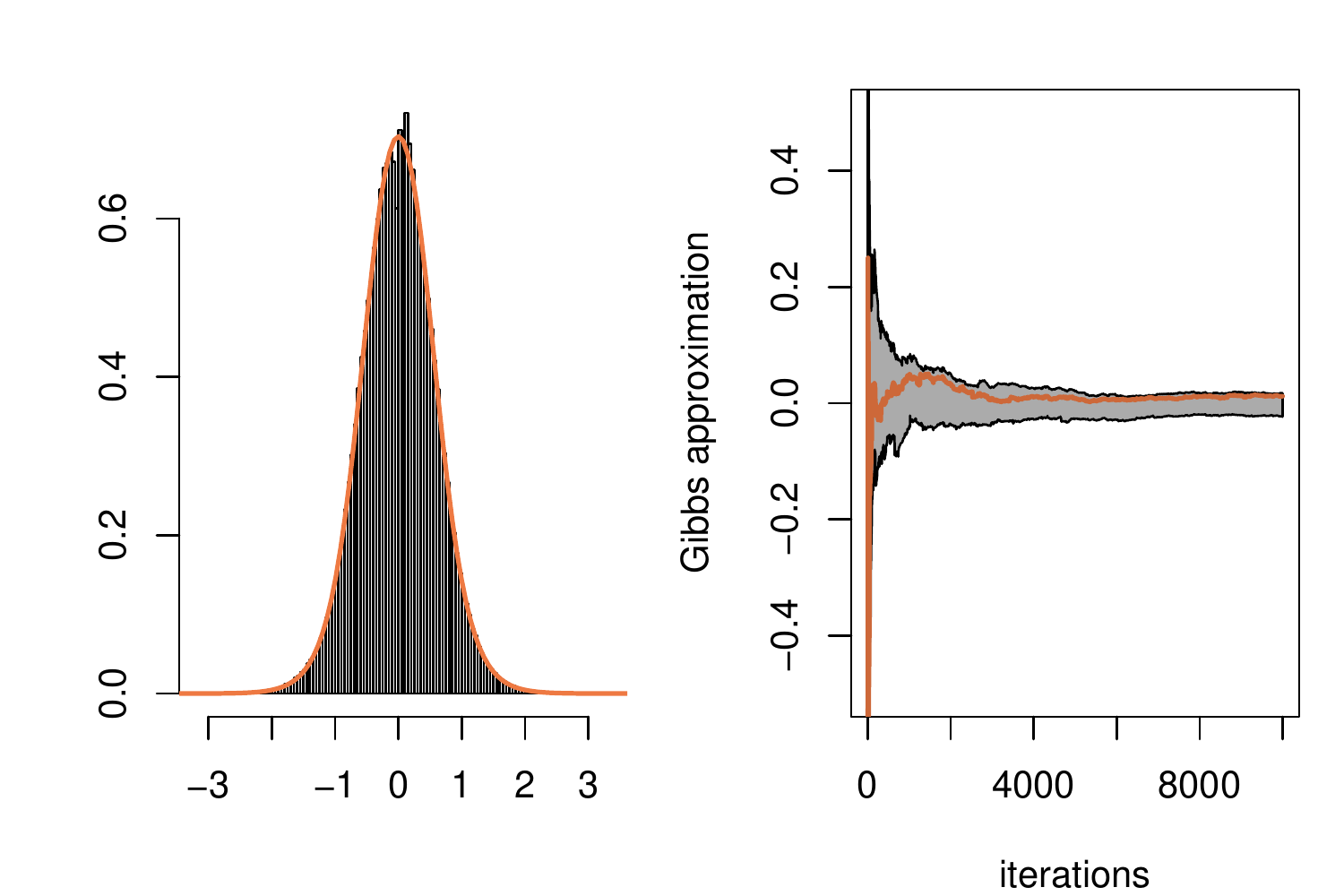}}
\caption{\label{fig:mcmcase}
{\em (left)} 
Gibbs sampling approximation to the distribution $f(x)\propto \exp(-x^2/2)/(1+x^2+x^4)$ against the true density;
{\em (right)}
range of convergence of the approximation to $\mathbb{E}_f[X^3]=0$ against
the number of iterations using 100 independent runs of the Gibbs sampler, along
with a single Gibbs run.
}
\end{figure}

For instance, if the density $f(x)\propto \exp(-x^2/2)/(1+x^2+x^4)$ is known up to a normalising
constant, $f$ is the marginal (in $x$) of the
joint distribution $g(x,u)\propto\exp(-x^2/2) \mathbb{I}(u(1+x^2+x^4)\le 1)$, when $u$ is restricted to $(0,1)$. The
corresponding slice sampler then consists in simulating 
$$
U|X=x\sim\mathcal{U}(0,1/(1+x^2+x^4))
$$
and
$$
X|U=u\sim\mathcal{N}(0,1)\mathbb{I}(1+x^2+x^4\le 1/u)\,,
$$
the later being a truncated normal distribution.
As shown by Figure \ref{fig:mcmcase},
the outcome of the resulting Gibbs sampler perfectly fits the target density, while the convergence of
the expectation of $X^3$ under $f$ has a behaviour quite comparable with the iid setting.

\begin{figure}
\centerline{\includegraphics[width=.45\textwidth]{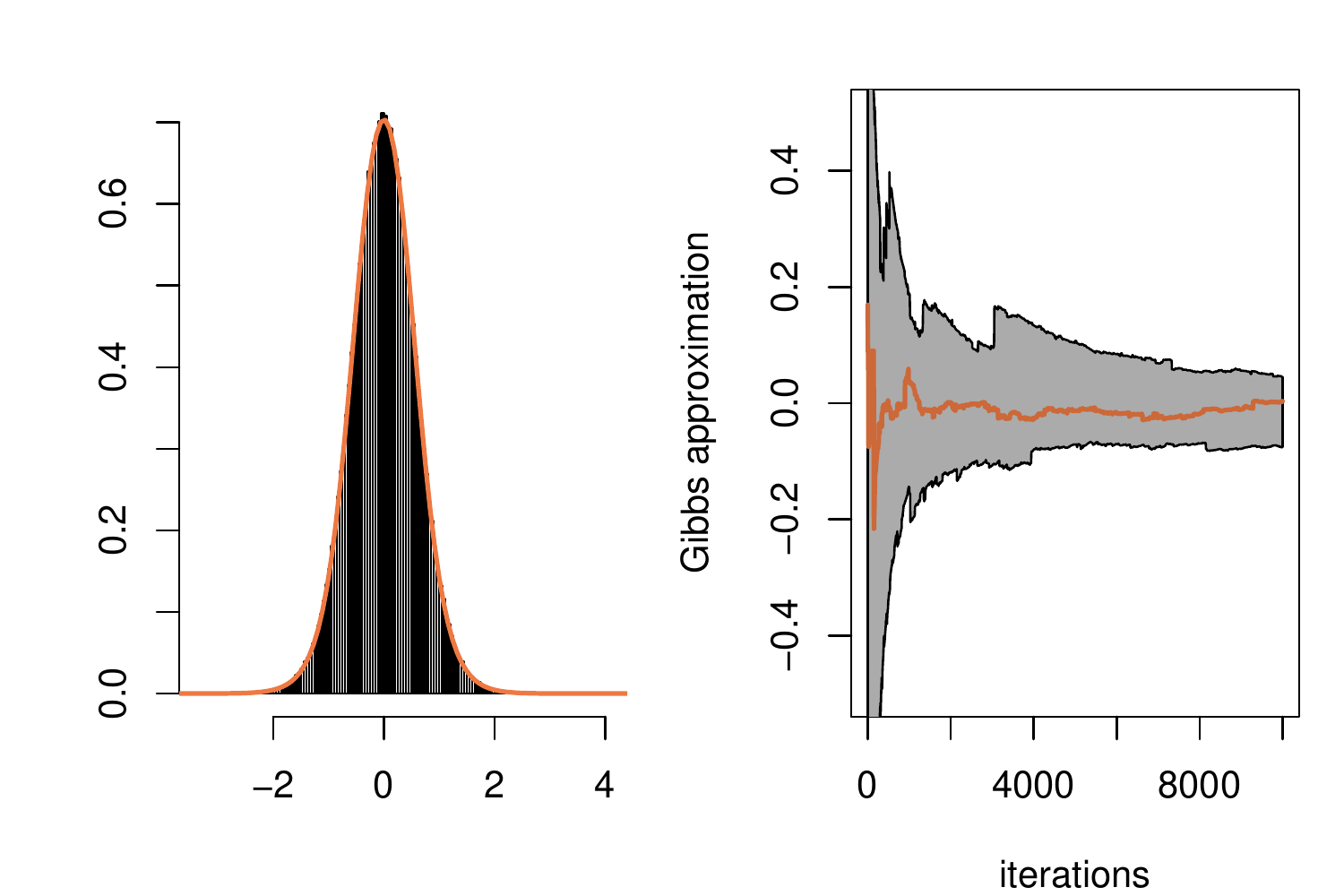}}
\caption{\label{fig:mcmet}
{\em (left)}
Random walk Metropolis--Hastings sampling approximation to the distribution $f(x)\propto \exp(-x^2/2)/(1+x^2+x^4)$ against the true density
for a scale of $1.2$ corresponding to an acceptance rate of $0.5$;
{\em (right)}
range of convergence of the approximation to $\mathbb{E}_f[X^3]=0$ against
the number of iterations using 100 independent runs of the Metropolis--Hastings sampler, along
with a single Metropolis--Hastings run.
}
\end{figure}

While the Gibbs sampler first appears as {\em the} natural solution to solve a simulation problem  
in complex models if only because it stems from the true target $f$, as exhibited by the widespread 
use of BUGS \cite{lunn:thomas:best:spiegelhalter:2000}, which mostly focus on this approach, 
the infinite variations offered by the Metropolis--Hastings schemes offer much more efficient solutions when the proposal 
$q(y|x)$ is appropriately chosen. The basic choice of a random walk proposal $q(y|x)$ being then a normal density centred
in $x$) can be improved by exploiting some features of the target as in Langevin algorithms (see \citealp{robert:casella:2004},
section 7.8.5) and Hamiltonian or hybrid alternatives \citep{duane:etal:1987,neal:1996} that build upon gradients. More recent proposals 
include particle learning about the target and sequential improvement of the proposal 
\citep{douc:guillin:marin:robert:2005,rosenthal:2007,andrieu:doucet:holenstein:2010}.
Figure \ref{fig:mcmet} reproduces Figure \ref{fig:mcmcase} for a random walk Metropolis--Hastings algorithm whose
scale is calibrated towards an acceptance rate of $0.5$. The range of the convergence paths is clearly wider than 
for the Gibbs sampler, but the fact that this is a generic algorithm applying to any target (instead of a specialised
version as for the Gibbs sampler) must be borne in mind.

Another major improvement generated by a statistical imperative is the development of variable dimension
generators that stemmed from Bayesian model choice requirements, the most important example being the
reversible jump algorithm in \cite{green:1995} which had a significant impact on the study of graphical 
models \citep{brooks:giudici:roberts:2003}.

\section*{Some uses of Monte Carlo in Statistics}

The impact of Monte Carlo methods on Statistics has not been truly felt until the early 1980's, with
the publication of \cite{rubinstein:1981} and \cite{ripley:1987}, but Monte Carlo methods have now
become invaluable in Statistics because they allow to address optimisation, integration and
exploration problems that would otherwise be unreachable. For instance, the calibration of many tests
and the derivation of their acceptance regions can only be achieved by simulation techniques.
While integration issues are often linked with the
Bayesian approach---since Bayes estimates are posterior expectations like
$$
\int h(\theta) \pi(\theta|x)\,\text{d}\theta
$$
and Bayes tests also involve integration, as mentioned earlier with the Bayes factors---,
and optimisation difficulties with the likelihood perspective, this classification is by no
way tight---as for instance when likelihoods involve unmanageable integrals---and all fields of Statistics, from design to econometrics, 
from genomics to psychometry and environmics, 
have now to rely on Monte Carlo approximations. A whole new range of statistical methodologies have entirely
integrated the simulation aspects. Examples include the bootstrap methodology \citep{efron:1982}, where multilevel resampling
is not conceivable without a computer, indirect inference \citep{gourieroux:monfort:renault:1993}, which construct a
pseudo-likelihood from simulations, MCEM \citep{cappe:moulines:2009}, where the E-step of the EM algorithm is replaced
with a Monte Carlo approximation, or the more recent approximated Bayesian computation (ABC) used in population genetics
\citep{beaumont:zhang:balding:2002}, where the likelihood is not manageable but the underlying model can be simulated
from.  

In the past fifteen years, the collection of real problems that Statistics can [afford to] handle has truly undergone a quantum leap.
Monte Carlo methods and in particular MCMC techniques have forever changed the emphasis from ``closed form" solutions to 
algorithmic ones, expanded our impact to solving ``real" applied problems while convincing scientists from other fields
that statistical solutions were indeed available, and led us into a world where ``exact" may mean ``simulated".
The size of the data sets and of the models currently handled thanks to those tools, 
for example in genomics or in climatology, is something that could
not have been conceived $60$ years ago, when Ulam and von Neumann invented the Monte Carlo method.

\small
\bibliographystyle{ims}
\bibliography{R09}

\end{document}